\documentclass[lettersize,journal]{IEEEtran}
\usepackage[flushleft]{threeparttable}

\usepackage{caption}

\usepackage{placeins}

\usepackage{enumitem}
\setlist[itemize,1]{label=$\times$}
\setlist[itemize,2]{label=$\checkmark$}
\setlist[itemize,3]{label=$\diamond$}
\setlist[itemize,4]{label=$\bullet$}
\usepackage{cite}
\usepackage[final]{hyperref}
\usepackage{listings}
\usepackage{xcolor}
\usepackage{pifont}
\usepackage{multirow}
\usepackage{caption}
\usepackage{tcolorbox}
\definecolor{codebg}{rgb}{0.95,0.95,0.95}
\definecolor{comment}{rgb}{0,0.6,0}
\definecolor{keyword}{rgb}{0,0,1}
\definecolor{string}{rgb}{0.58,0,0.82}

\lstdefinelanguage{Solidity}{
  morekeywords={pragma, solidity, contract, address, public, uint, require, mapping, bool, modifier, onlyOwner, string, bytes},
  sensitive=true,
  morecomment=[l]{//},
  morecomment=[s]{/*}{*/},
  morestring=[b]"
}

\usepackage{circledsteps}

\usepackage{amsmath,amsfonts}
\usepackage{algorithmic}
\usepackage{algorithm}
\usepackage{array}
\usepackage[caption=false,font=normalsize,labelfont=sf,textfont=sf]{subfig}
\usepackage{textcomp}
\usepackage{stfloats}
\usepackage{url}
\usepackage{verbatim}
\usepackage{graphicx}
\usepackage{cite}

\usepackage{colortbl}
\definecolor{LightCyan}{rgb}{0.88,1,1}
\definecolor{LightViolet}{RGB}{244,225,252}
\definecolor{LightYellow}{RGB}{255,255,204}
\definecolor{LightGreen}{RGB}{204,255,229}
\definecolor{LightOrinoco}{RGB}{233,250,207}
\definecolor{LightBlue}{RGB}{210,224,251}
\definecolor{LightRed}{RGB}{255,222,222}
\definecolor{LightOrange}{RGB}{255,227,187}
\definecolor{LightCream}{RGB}{252,239,230}

\begin{document}

\title{Quantum-Resilient Secure Protocol for Decentralized Edge-IoT Peer-to-Peer Energy Trading Systems}

\author{Ali Shahidinejad, Jamshid Aghaei, and Mahdi Jalili
\thanks{Manuscript received x x, 2026; revised x x, 2026; accepted x xx, xxxx. (\emph{Corresponding author: Ali Shahidinejad})}
\thanks{Ali Shahidinejad and Jamshid Aghaei are with the School of Engineering and Technology, Central Queensland University, Rockhampton, QLD , 4701, Australia, Email: a.shahidinejad@cqu.edu.au and j.aghaei@cqu.edu.au. Mahdi Jalili is with the Department of Electrical and Electronics Engineering, Royal Melbourne Institute of Technology University, Melbourne, 3000, VIC, Australia, Email: mahdi.jalili@rmit.edu.au}}

\markboth{}%
{Shell \MakeLowercase{\textit{et al.}}: A Novel Security Protocol for Quantum-Era Decentralized Edge-IoT P2P Energy Trading}


\maketitle

\begin{abstract}
The rise of quantum computing presents substantial challenges to the security of Peer-to-Peer (P2P) energy trading systems. This paper introduces an innovative mutual authenticated session key agreement protocol aimed at protecting edge-IoT P2P energy trading communications in the quantum era. Each peer is an IoT-enabled smart metering device that is registered by an edge registration authority and acts as both producer and consumer. The proposed protocol incorporates CRYSTALS-Kyber's quantum-resistant cryptographic suite to guarantee the authenticity and confidentiality of negotiated session keys in the presence of quantum threats. Furthermore, it leverages the Hedera Hashgraph public blockchain, which is recognized for its quantum-resilient design, to ensure transparency and support efficient revocation mechanisms. Extensive comparative analyses have been conducted to demonstrate the protocol's superiority over existing alternatives, including both traditional schemes based on classical cryptography and emerging early-stage post-quantum approaches, in terms of performance and security features. By combining robust security and high efficiency, the proposed protocol fosters the development of resilient decentralized energy markets, paving the way for secure and sustainable energy distribution in the future.
\end{abstract}


\section{Introduction}
\IEEEPARstart{T}{he} rapid advancement of quantum computing technology is poised to revolutionize various fields, including cryptography, data security, and decentralized systems. While quantum computing holds immense potential for solving complex problems, it also introduces significant security challenges, particularly for decentralized Peer-to-Peer (P2P) energy trading systems \cite{capellaro2025application}. In such systems, households or entities can directly trade surplus energy, for example solar power, with one another without relying on a central intermediary. These systems, which enable direct energy transactions between Internet of Things (IoT)-enabled producer and consumer smart meters, rely heavily on cryptographic protocols to ensure secure and trustworthy operations.

As quantum computers become more powerful, they pose a serious threat to widely used cryptographic algorithms such as RSA and Elliptic Curve Cryptography (ECC), as they can efficiently solve the underlying mathematical problems that ensure their security. This vulnerability necessitates the development of new security protocols that can withstand quantum attacks and protect the integrity, confidentiality, and authenticity of transactions. Addressing this critical issue, our research focuses on designing a novel decentralized P2P session key agreement protocol with quantum-resilient security based on post-quantum cryptographic mechanisms.

Our security scheme leverages CRYSTALS-Kyber's \cite{bos2018crystals} cryptographic suite and the Hedera blockchain platform, both of which are designed to be quantum-resilient. By integrating advanced blockchain technology with quantum-safe algorithms, our protocol not only provides a robust solution to emerging quantum threats but also ensures key security properties, including revocability, fault tolerance, transparency, and decentralization. We conduct extensive analyses demonstrating its capability to maintain secure and efficient energy trading even in the presence of quantum-capable adversaries.

Our findings contribute to ongoing efforts to develop resilient decentralized energy markets \cite{sampath2021peer}. By ensuring the security and sustainability of P2P energy trading, this work supports a future in which energy distribution is both efficient and secure against evolving cyber threats. This research highlights the importance of proactive security design in safeguarding critical infrastructure as we transition into the quantum era.

\subsection{Related Work}
This section reviews top-related smart energy protocols, summarized in Table~\ref{tab:related}.

\subsubsection{Blockchain-Based Quantum-Secure Protocols}
This group comprises security protocols that are both quantum-secure and built on blockchain technology. The proposed protocols in this group \cite{shekhawat2024quantum2, verma2022sp2p} are the most preferable ones since they can provide both the revocation option and quantum-resistance. Although some blockchain-based quantum-secure protocols have been already proposed, the authors did not utilize quantum-resistant platforms to achieve blockchain-level quantum security or few used public blockchain platforms to achieve full transparency. More critically, their employed quantum suites have not been approved by the National Institute of Standards and Technology (NIST) \cite{alagic2025status}.

\subsubsection{Quantum-Secure Protocols Without Blockchain}
This group comprises security protocols that are just quantum-secure. The proposed protocols in \cite{shekhawat2024quantum1,lu2021quantum} are in this group. These protocols cannot easily provide the revocation of smart metering devices. Nevertheless, as they are quantum-secure, they are more preferable than the classical ones.

\subsubsection{Blockchain-Based Classical Protocols}
This group comprises security protocols that are built on blockchain technology. The proposed protocols by \cite{wang2019blockchain,shahidinejad2024highly} are in this group. These protocols are vulnerable to quantum attacks; but they are better than classical ones because, by the blockchain integration, they can offer features like revocation option.

\subsubsection{Classical Protocols Without Blockchain}
This group comprises security protocols that are not quantum-secure neither built on blockchain technology. Most prior works \cite{abbasinezhad2018anonymous, abbasinezhad2018design, qi2020two, baruah2021authenticated, chaudhry2021privacy, hu2022provably, chaudhry2022sg} are in this group. Although sometimes they can offer a better performance level, they cannot easily provide transparency/revocability neither can resist quantum computers.

\begin{table}[!t]
\scriptsize
\renewcommand{\arraystretch}{1.1}
\setlength{\tabcolsep}{0.07em}
\caption{Comparison of Key Features}
\label{tab:related}
\centering
\begin{threeparttable}
\begin{tabular}{*{14}{c @{\hspace{0.2 cm}} c @{\hspace{0.2 cm}} c @{\hspace{0.2 cm}} c @{\hspace{0.2 cm}} c @{\hspace{0.2 cm}} c @{\hspace{0.2 cm}} c @{\hspace{0.2 cm}} c @{\hspace{0.2 cm}} c @{\hspace{0.2 cm}} c @{\hspace{0.2 cm}} c @{\hspace{0.2 cm}} c @{\hspace{0.2 cm}} c} }\hline\hline

Scheme  &F1    &F2    &F3    &F4    &F5    &F6    &F7    &F8    &F9    &F10   &F11   &F12 \\\hline
\rowcolor{LightOrinoco}\multicolumn{13}{c}{\footnotesize{Blockchain-Based Quantum-Secure Protocols}}\\\hline

Ours
&CRYSTALS-Kyber
&eCK
&Hedera
&B
&\ding{51} 
&\ding{51}
&\ding{51}
&\ding{51}
&\ding{51}
&\ding{51}
&\ding{51}
&\ding{51}\\\hline
\cite{verma2022sp2p} &Lattice/ISIS &NS &NS &\ding{55} &\ding{55} &\ding{55} &\ding{55} &\ding{51} &\ding{51} &\ding{55} &\ding{55} &\ding{55}\\\hline

\cite{shekhawat2024quantum2} &Lattice/ISIS &UD &HLF &\ding{51} &\ding{51} &\ding{51} &\ding{51} &\ding{51} &\ding{51} &\ding{55} &\ding{55} &\ding{55}\\\hline

\rowcolor{LightRed}\multicolumn{13}{c}{\footnotesize{Quantum-Secure Protocols Without Blockchain}}\\\hline

\cite{lu2021quantum} &HB+/LPN &CK &--- &B &\ding{51} &\ding{55} &\ding{55} &\ding{51} &\ding{55} &\ding{55} &--- &\ding{55}\\\hline

\cite{shekhawat2024quantum1} &Lattice/ISIS &UD &--- &\ding{51} &\ding{51} &\ding{51} &\ding{55} &\ding{55} &\ding{55} &\ding{55} &--- &\ding{55}\\\hline

\rowcolor{LightBlue}\multicolumn{13}{c}{\footnotesize{Blockchain-Based Classical Protocols}}\\\hline 

\cite{wang2019blockchain} &ECC &UD &HLF &\ding{51} &\ding{51} &\ding{55} &\ding{51} &\ding{51} &\ding{51} &\ding{55} &\ding{55} &---\\\hline
\cite{shahidinejad2024highly} &ECC &CK &HLF &B &\ding{51} &\ding{55} &\ding{51} &\ding{51} &\ding{51} &\ding{55} &\ding{55} &---\\\hline

\rowcolor{LightOrange}\multicolumn{13}{c}{\footnotesize{Classical Protocols Without Blockchain}}\\\hline

\cite{abbasinezhad2018anonymous} &ECC &CK &--- &B &\ding{51} &\ding{55} &\ding{55} &\ding{51} &\ding{55} &\ding{55} &--- &---\\\hline
\cite{abbasinezhad2018design} &ECC &CK &--- &\ding{51} &\ding{51} &\ding{55} &\ding{55} &\ding{55} &\ding{55} &\ding{55} &--- &---\\\hline
\cite{qi2020two} &ECC &CK &--- &\ding{51} &\ding{55} &\ding{55} &\ding{55} &\ding{51} &\ding{55} &\ding{55} &--- &---\\\hline
\cite{baruah2021authenticated} &ECC &UD &--- &B &\ding{51} &\ding{55} &\ding{55} &\ding{51} &\ding{55} &\ding{55} &--- &---\\\hline
\cite{chaudhry2021privacy} &ECC &eCK &--- &B &\ding{51} &\ding{55} &\ding{55} &\ding{51} &\ding{55} &\ding{55} &--- &---\\\hline
\cite{hu2022provably} &ECC &CK &--- &B &\ding{51} &\ding{55} &\ding{55} &\ding{51} &\ding{55} &\ding{55} &--- &---\\\hline
\cite{chaudhry2022sg} &ECC &eCK &--- &B &\ding{51} &\ding{55} &\ding{55} &\ding{51} &\ding{55} &\ding{55} &--- &---\\\hline\hline

\end{tabular}
F: Feature; F1: Cryptosystem; F2: Attack model; F3: Blockchain platform; F4: Formal verification by software or formal proof or both; F5: Hardware performance evaluation; F6: Mutual session-key verification; F7: Conditional anonymity; F8: Unlinkability; F9: Revocability; F10: Full transparency; F11: Quantum-proof blockchain; F12: NIST-approved quantum suite. HLF: Hyperledger Fabric; ISIS: Inhomogeneous Small Integer Solution; HB: Hopper-Blum; LPN: Learning Parity with Noise; NS: Not Specified; UD: User Defined; CK: Canetti-Krawczyk; eCK: extended CK; ECC: Elliptic Curve Cryptography; B: Both; P: Partial; ---: Not applicable; \ding{51}: Supported; \ding{55}: Not supported.
\end{threeparttable}
\end{table}

\subsection{Motivation and Contribution}
As reviewed earlier in the previous subsection, none of the existing works can simultaneously provide the desired security properties, such as transparency (F10), revocability (F9), and quantum-level security (F11, F12), as summarized in Table~\ref{tab:related}. This limitation highlights a critical gap in current solutions. Therefore, this gap has motivated us to propose a novel protocol that offers the following key contributions.

\begin{itemize}[label={$\checkmark$}]
\item The proposed protocol leverages CRYSTALS-Kyber as one of the NIST-approved quantum-secure cryptographic suite to safeguard P2P communications of smart metering devices against quantum threats.
\item The proposed protocol utilizes the distributed ledger of blockchain as a trustful platform for storing/retrieving public keys of SMDs. Storing public keys in the ledger simplifies public key revocation.
\item The proposed protocol leverages Hedera Hashgraph public blockchain platform to achieve transparency and resilience to quantum attacks. By the way, the Hedera Hashgraph is among the few quantum-resistance ones.
\item The proposed protocol is both efficient and secure; it resists advanced attacks like Key Compromise Impersonation (KCI) and offers certain features like conditional anonymity and Perfect Forward Secrecy (PFS). More significantly, compared to most related protocols, it puts least overhead on computationally-limited smart meters. 
\item In the proposed protocol, instead of one central trusted registration authority, several semi-trusted edge registration authorities perform the task enhancing load-balancing, fault-tolerance, and decentralization.
\end{itemize}

\section{System and Attack Models}

\subsection{System Model}\label{Section2.1}
The system model of the proposed protocol consists of multiple semi-trusted Edge Registration Authorities (ERAs) and Smart Metering Devices (SMDs). Note that an ERA can be a third-party entity responsible for manufacturing or provisioning SMDs. The term semi-trusted indicates that, while ERAs are expected to follow the protocol during system initialization and registration, they are not fully trusted against compromise. In particular, if an ERA is compromised, the adversary may gain access to registration-related data or attempt to register malicious devices; however, the overall system does not collapse. This is because ERAs are not involved in real-time authentication between SMDs. The use of decentralized blockchain storage and device-generated key pairs ensures that peer-to-peer energy trading sessions remain secure, thereby limiting the impact of a compromised ERA.

\begin{figure}[!t]
\scriptsize
\centering
\includegraphics[width=2.9in]{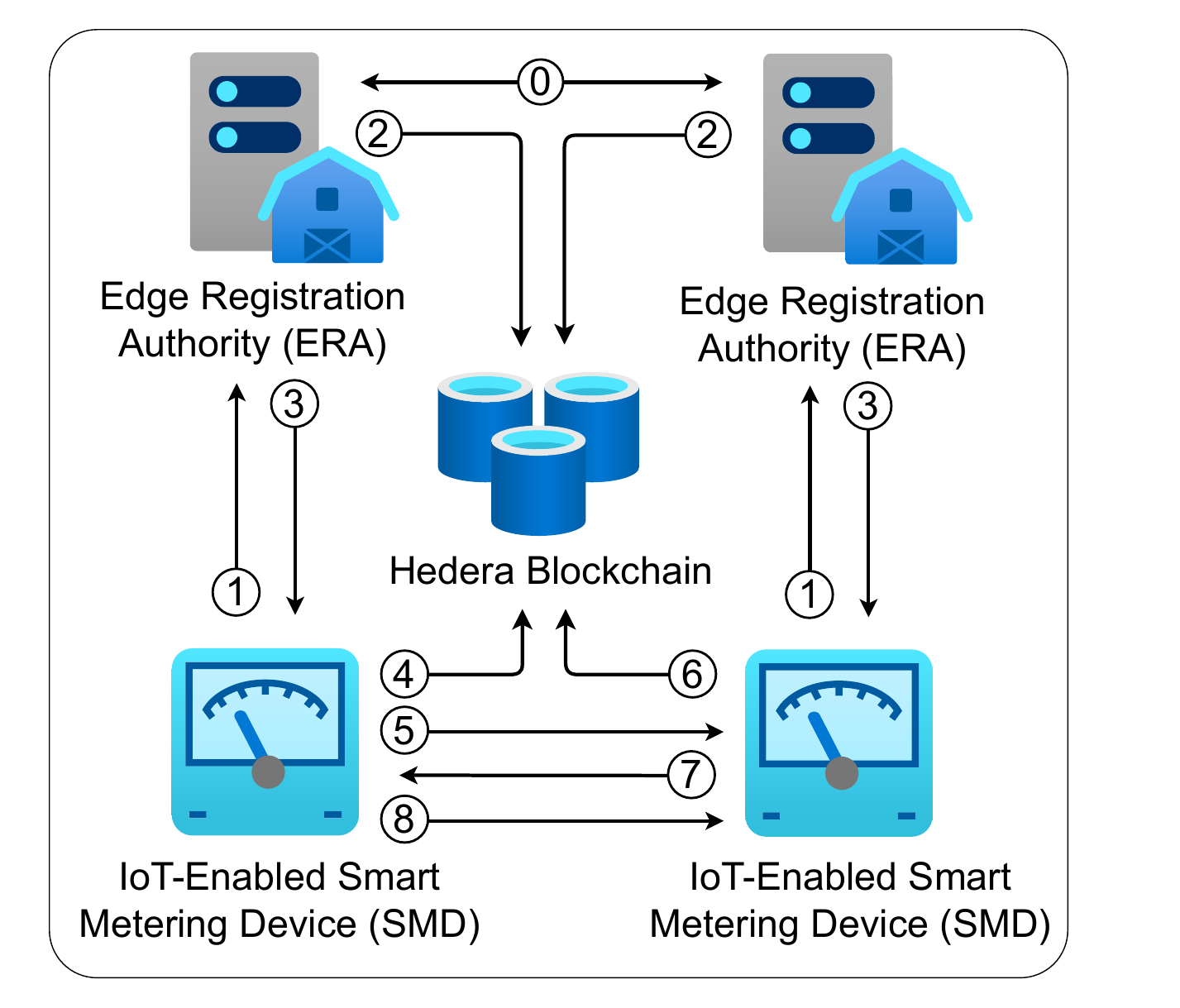}
\caption{System model and communications of protocol participants. Descriptions are presented in Subsection~\ref{Section2.1}.}
\label{fig00}
\end{figure}

There exists a Hedera blockchain in the middle that acts as a reliable storage/retrival place for SMDs' public keys. Fig.~\ref{fig00} depicts the communication model and steps. The aim of the proposed protocol is to perform a mutual P2P authentication and session key establishment between two SMDs, which are described according to the following steps. \Circled{0} ERAs perform system initialization by choosing the CRYSTALS-Kyber's cryptographic suite and Hedera blockchain. Further, they develop and deploy a secure Solidity smart contract for storing/retrieving public keys to/from the blockchain. \Circled{1} After getting the system parameters from the ERA, SMD generates its private/public key pair. The SMD keeps the private key secure (using either tamper-proof or physically unclonable function mechanism) and submits a registration request to ERA containing its public key. \Circled{2} The ERA then generates an encrypted identifier for SMD and stores public key and verification token in the Hedera blockchain. \Circled{3} ERA returns encrypted identifier to SMD as the registration response. \Circled{4} Initiator SMD reads public key of target SMD, which intends to share a key with, from Hedera blockchain and \Circled{5} sends an encrypted authentication request using the retrieved public key. \Circled{6} After checking time and integrity of message, the target SMD generates session key and reads the public key of the initiator SMD from the blockchain. \Circled{7} Target SMD responds back the initiator SMD with an encrypted authentication response using the retrieved public key. \Circled{8} Initiator SMD computes the session key and sends an acknowledgement message to the target SMD. The target SMD would finally accept the session key if the acknowledgement check is successful.

\subsection{Attack Model} 
The extended Canetti–Krawczyk (eCK) adversary model \cite{lamacchia2007stronger} is employed to evaluate the security of the proposed P2P energy trading protocol. The eCK model is widely regarded as one of the most stringent and comprehensive security models for authenticated key agreement, as it captures a broad range of realistic and powerful adversarial capabilities. It assumes an adversary capable of intercepting, modifying, and injecting messages during exchanges between authorized entities, with the objective of impersonating a legitimate participant even in the presence of certain key compromises.

This model enables a rigorous assessment of the protocol’s robustness by considering scenarios such as exposure of long-term keys and ephemeral secrets. Consequently, it provides a strong indication of the protocol’s resilience against advanced attacks, which is particularly important in the context of emerging quantum computing threats. Table~\ref{Queries} outlines the distinct capabilities of the adversary $Att_{qe}^{ck}$, represented as query types. These include intercepting communication messages, accessing long-term cryptographic keys, and revealing participant identities or established session keys.
\begin{table}[!htbp]
\setlength{\tabcolsep}{0.1em}
\scriptsize
\caption{Queries}   
\label{Queries}
\centering
\begin{threeparttable}
\renewcommand{\arraystretch}{1}
\begin{tabular}{ll}
\hline\hline
\rowcolor{LightOrinoco} Query & Explanation \\
\hline
$SHA3(Inp)$                   &$Att_{qe}^{ck}$ gets SHA3 hashed value of $Inp$.\\
$Execute(SMD_a, SMD_b)$       &$Att_{qe}^{ck}$ gets messages between $SMD_a$ and $SMD_b$.\\
$Send(SMD_a, Req)$            &$Att_{qe}^{ck}$ sends $Req$ to $SMD_a$ and receives response.\\
$Send(SMD_b, Req)$            &$Att_{qe}^{ck}$ sends $Req$ to $SMD_b$ and receives response.\\
$Corrupt(SMD_a)$              &$Att_{qe}^{ck}$ gets long-term secrets of $SMD_a$.\\
$Corrupt(SMD_b)$              &$Att_{qe}^{ck}$ gets long-term secrets of $SMD_b$.\\
$TestID(SMD_a | SMD_b)$       &$Att_{qe}^{ck}$ gets ID of $SMD_a$ or $SMD_b$.\\
$TestSSK(SMD_a, SMD_b)$       &$Att_{qe}^{ck}$ gets SSK between $SMD_a$ and $SMD_b$.\\
\hline\hline
\end{tabular}
Note: $TestID(SMD_a | SMD_b)$ and $TestSSK(SMD_a, SMD_b)$ can be issued just once. $Req$: Request message; $Inp$: Input value; $SMD$: Smart Metering Device; $SMD_a$: $a^{th}$ instance of SMD communicating with $SMD_b$, $SMD_b$: $b^{th}$ instance of SMD communicating with $SMD_a$; $Att_{qe}^{ck}$: Quantum-Enabled CK attacker.
\end{threeparttable}
\end{table}

\section{Conceptual Framework}

\subsection{CRYSTALS-Kyber}
CRYSTALS-Kyber \cite{alagic2025status} is an advanced cryptographic algorithm developed to tackle the challenges introduced by the emergence of quantum computing. Part of the CRYSTALS (Cryptographic Suite for Algebraic Lattices) group, it functions as a key encapsulation mechanism within the realm of post-quantum cryptography. It is rooted in lattice-based cryptography, widely regarded as highly secure against attacks from both classical and quantum systems. CRYSTALS-Kyber is praised for its effectiveness, delivering rapid key creation, encryption, and decryption, making it ideal for diverse applications, including secure communications of IoT devices. Its significance is emphasized by its selection in the NIST post-quantum cryptography standardization initiative, underlining its capability to protect critical data in a quantum-empowered future. Additionally, CRYSTALS-Kyber is immensely valuable in fortifying decentralized P2P energy trading frameworks against quantum-driven risks. By safeguarding communication pathways between energy providers and consumers, it ensures the security and confidentiality of energy trades, promoting trustworthiness in modern energy systems.

In this paper, we utilize three Kyber's algorithms, namely $KKG_{CPAKE}()$, which represents Chosen Plaintext Attack Public Key Encryption (CPAPKE) Kyber Key Generator Function, $KE_{CPAKE}(PK, PT, r)$, which represents CPAPKE Kyber Encryption Function, and $KD_{CPAKE}(SK, CT)$, which represents CPAPKE Kyber Decryption Function, where $B$, $SK$, $PK$, $PT$, $CT$, and $r$, respectively, denote byte, secret key, public key, plain-text, cipher-text, and random number. The relation of these three is defined formally as Equation~(\ref{0}).

\begin{equation}{}\label{0}
\begin{split}
(PK, SK) = KKG_{CPAKE}() \mid SK \in \mathcal{B}^{768}, PK \in \mathcal{B}^{800}\\
CT = KE_{CPAKE}(PK, PT, r) \mid PT \in \mathcal{B}^{32}, r \in \mathcal{B}^{32} \quad\\
PT = KD_{CPAKE}(SK, CT) \mid CT \in \mathcal{B}^{768} \qquad \quad
\end{split}
\end{equation}

\subsection{Hedera Hashgraph Public Blockchain}
Hedera Hashgraph emerges as a state-of-the-art distributed ledger technology, showcasing exceptional velocity, robustness, and effectiveness. Diverging from conventional blockchains, it utilizes an innovative hashgraph consensus mechanism, enabling it to handle over 10,000 transactions per second with minimal delay. Its asynchronous Byzantine Fault Tolerance (aBFT) guarantees unparalleled security, making it impervious to malicious threats. Furthermore, Hedera is engineered to be quantum-proof, preserving its cryptographic stability against potential quantum computing advancements. The platform is highly energy-efficient, requiring significantly less power compared to traditional blockchains. This eco-friendliness and scalability position Hedera as an optimal choice for decentralized P2P energy exchange, where secure, instantaneous transactions between energy producers and consumers are critical. Governed by a consortium of leading global organizations, Hedera ensures reliability and decentralization while offering consistent transaction costs. These attributes render it a perfect fit for decentralized infrastructures and inventive solutions like P2P energy trading.

\begin{figure}[ht!]
    \centering
    \tiny
    \begin{lstlisting}[numbers=none]
// SPDX-License-Identifier: MIT
pragma solidity ^0.8.0;

// SecureStorage contract for securely managing records with ownership controls
contract SecureStorage {
    // Mapping to store owners. An address is true if it's an authorized owner.
    mapping(address => bool) public owners;

    // Struct to store records with essential fields
    struct Record {
        bytes12 EID;        // Unique identifier for the record (12 bytes)
        bytes PK;           // Public key (PK must be exactly 800 bytes)
        bytes32 V;          // Verification field (32 bytes)
        bool revoked;       // Boolean flag to indicate if the record is revoked
    }

    // Mapping of unique EID to Record structure
    mapping(bytes12 => Record) private records;

    // Modifier to restrict access to only authorized owners
    modifier onlyOwner() {
        require(owners[msg.sender], "Not an authorized owner"); // Check if the sender is an owner
        _;
    }

    // Constructor that sets the deployer as the initial owner
    constructor() {
        owners[msg.sender] = true; // The contract deployer is the first owner
    }

    // Function to add a new owner
    function addOwner(address newOwner) public onlyOwner {
        require(newOwner != address(0), "Invalid address"); // Ensure the address is valid
        owners[newOwner] = true; // Grant ownership to the specified address
    }

    // Function to remove an existing owner
    function removeOwner(address ownerToRemove) public onlyOwner {
        require(ownerToRemove != msg.sender, "Owners cannot remove themselves"); // Prevent self-removal
        owners[ownerToRemove] = false; // Revoke ownership
    }

    // Function to store a new record
    function storeRecord(bytes12 _EID, bytes memory _PK, bytes32 _V, bool _revoked) public onlyOwner {
        require(_PK.length == 800, "PK must be 800 bytes long"); // Validate PK length
        records[_EID] = Record(_EID, _PK, _V, _revoked); // Store the record in the mapping
    }

    // Function to retrieve a record by its EID
    function retrieveRecord(bytes12 _EID) public view returns (bytes12, bytes memory, bytes32, bool) {
        Record memory record = records[_EID]; // Fetch the record from the mapping
        return (record.EID, record.PK, record.V, record.revoked); // Return all fields of the record
    }

    // Function to mark a record as revoked
    function revokeRecord(bytes12 _EID) public onlyOwner {
        require(records[_EID].EID != 0, "Record does not exist"); // Ensure the record exists
        records[_EID].revoked = true; // Set the revoked flag to true
    }

    // Function to check if a record is revoked
    function isRevoked(bytes12 _EID) public view returns (bool) {
        return records[_EID].revoked; // Return the revoked status of the record
    }
}

    \end{lstlisting}
    \caption{Solidity contract for secure storage and retrieval of smart metering device information in the Hedera blockchain.}
    \label{fig:solidity-smart-contract}
\end{figure}

\section{Proposed Protocol}
``Initial Setup,'' ``Smart Metering Device Registration,'' ``Smart Metering Device P2P Mutual Authentication,'' and ``Smart Metering Device Revocation'' are the four phases of the proposed protocol. The general overview of the first three phases has been depicted in Fig.~\ref{fig00} and summarized in Subsection~\ref{Section2.1}. Details of each are elaborated as follows.

\begin{table}[!htbp]
\setlength{\tabcolsep}{1em}
\caption{Notations} 
\scriptsize
\label{notations}
\centering
\renewcommand{\arraystretch}{1}
\begin{tabular}{ll}
\hline\hline
\rowcolor{LightOrinoco} Term & Meaning\\
\hline
$SMD_a/SMD_b$              &$a^{th}/b^{th}$ Smart Metering Device\\
$ERA_c/ERA_d$              &$c^{th}/d^{th}$ Edge Registration Authority\\
$SK/PK$                    &Secret Key/Public Key \\
$SSK$                      &Shared Session Key \\
$ID/EID$                   &Identifier/Encrypted Identifier \\
$A/V$                      &Acknowledgement/Verification \\
$K/C$                      &Key/Confidential \\
$E/R$                      &Encrypted/Random \\
$H_{SHA256}/H_{SHA384}$    &SHA256/SHA384 Hashing Function \\
$SE^{AES256}_{K}$          &AES256 Symmetric Encryption Using Key $K$\\
$SD^{AES256}_{K}$          &AES256 Symmetric Decryption Using Key $K$\\
$CPAPKE$                   &Chosen Plaintext Attack Public Key Encryption \\
$PT/CT$                    &Plain-text/Cipher-text \\
$KKG_{CPAKE}()$            &CPAPKE Kyber Key Generator Function \\
$KE_{CPAKE}(PK, PT, r)$    &CPAPKE Kyber Encryption Function\\
$KD_{CPAKE}(SK, CT)$       &CPAPKE Kyber Decryption Function\\
\hline\hline
\end{tabular}
\end{table}

\subsection{Initial Setup}
In this phase, semi-trusted Edge Registration Authorities (ERAs) set system parameters $Params$ that includes SHA256, SHA384, AES256, $KKG_{CPAKE}()$, $KE_{CPAKE}(PK, PT, r)$, and $KD_{CPAKE}(SK, CT)$ algorithms. They further develop the presented solidity smart contract in Fig.~\ref{fig:solidity-smart-contract} and deploy it to the public Hedera blockchain network. It is worth clarifying that only ERAs, which are owners, are allowed to modify the inserted records and others can only read the inserted records. This modification can include inserting a new record or revoke an existing one. 

\subsection{Smart Metering Device Registration}
After receiving system parameters $Params$, $a^{th}$ Smart Metering Device (SMD), denoted as $SMD_a$, generates its secret/public key pair by employing the CPAPKE Kyber Key Generator Function as $(PK_{SMD_a}, SK_{SMD_a}) = KKG_{CPAKE}()$. Next, it generates random $v_{SMD_a}$, computes verification value $V_{SMD_a} = H_{SHA256}(v_{SMD_a})$, and submits registration request message $\{ID_{SMD_a}, PK_{SMD_a}, V_{SMD_a}\}$ to $c^{th}$ Edge Registration Authority (ERA), denoted as $ERA_c$. Here we assume that $ERA_c$ is the manufacturer of $SMD_a$ or a registration authority in the vicinity of the owner of $SMD_a$. For $b^{th}$ SMD, denoted as $SMD_b$, the registration authority could be $d^{th}$ ERA, denoted as $ERA_d$.

After receiving the request, $ERA_c$ checks whether or not $SMD_a$ is within the registered ones and if not, generates random $x_{ERA_c}$ and encrypted identifier $EID_{SMD_a} = SE^{AES256}_{SK_{ERA_c}}(ID_{SMD_a}, x_{ERA_c})$, sets $IsRevoked_a = False$, stores $(EID_{SMD_a}, PK_{SMD_a}, V_{SMD_a}, IsRevoked_a)$ in the Hedera blockchain, and returns back $\{EID_{SMD_a}\}$ to $SMD_a$. $ERA_c$ calls the ``storeRecord'' function of the presented smart contract in Fig.~\ref{fig:solidity-smart-contract} to store the record. 

\subsection{Smart Metering Device P2P Mutual Authentication}
The process is embarked by $SMD_a$, as the initiator SMD, by generating session-random key pairs $(X_{SMD_a}, x_{SMD_a}) = KKG_{CPAKE}()$ and random $k_{SMD_a}, \allowbreak y_{SMD_a}, \allowbreak r_{SMD_a}$ and computing key $K_{SMD_a} = H_{SHA256}(k_{SMD_a})$. $SMD_a$ then encrypts required values for authentication and session key agreement as $E_{SMD_a} = SE^{AES256}_{K_{SMD_a}}(EID_{SMD_a}, \allowbreak X_{SMD_a}, \allowbreak y_{SMD_a}, \allowbreak v_{SMD_a})$. Next, by calling the ``retrieveRecord'' function of the presented smart contract in Fig.~\ref{fig:solidity-smart-contract}, it retrieves the information of its target SMD, such as $PK_{SMD_b}, \allowbreak V_{SMD_b}, \allowbreak IsRevoked_b$, from the Hedera blockchain. If $IsRevoked_b \stackrel{?}{=} False$, $SMD_a$ generates a 32-byte random $R_{SMD_a} = H_{SHA256}(r_{SMD_a})$ and using the retrieved public key $PK_{SMD_b}$ and CPAPKE Kyber Encryption Function, encrypts key $K_{SMD_a}$ as $C_{SMD_a} = KE_{CPAKE}(PK_{SMD_b}, \allowbreak K_{SMD_a}, \allowbreak R_{SMD_a})$. Doing so, only $SMD_b$ can decrypt $C_{SMD_a}$ and get access to symmetric key $K_{SMD_a}$. Finally, it computes integrity value $H_{SMD_a} = H_{SHA384}(EID_{SMD_a}, \allowbreak X_{SMD_a}, \allowbreak y_{SMD_a}, \allowbreak v_{SMD_a}, \allowbreak T_{SMD_a})$ and submits its request as $\{E_{SMD_a}, \allowbreak C_{SMD_a}, \allowbreak H_{SMD_a}, \allowbreak T_{SMD_a}\}$.

After receiving request and checking the time validity, using its secret key and CPAPKE Kyber Decryption Function, $SMD_b$ gets key $K_{SMD_a} = KD_{CPAKE}(SK_{SMD_b}, C_{SMD_a})$ and decrypts $E_{SMD_a}$ as $(EID_{SMD_a}, X_{SMD_a}, y_{SMD_a}, v_{SMD_a}) = SD^{AES256}_{K_{SMD_a}}(E_{SMD_a})$. Next, it checks $H_{SHA384}(EID_{SMD_a}, \allowbreak X_{SMD_a}, \allowbreak y_{SMD_a}, \allowbreak v_{SMD_a}, \allowbreak T_{SMD_a}) \stackrel{?}{=} H_{SMD_a}$ and if so, it reads $PK_{SMD_a}, \allowbreak V_{SMD_a}, IsRevoked_a$ from the Hedera blockchain by calling the ``retrieveRecord'' function of the presented smart contract in Fig.~\ref{fig:solidity-smart-contract}. It then checks $H_{SHA256}(v_{SMD_a}) \stackrel{?}{=} V_{SMD_a}$ and $IsRevoked_a \stackrel{?}{=} False$ and if so, it generates random $k_{SMD_b}, x_{SMD_b}, y_{SMD_b}, r_{SMD_b}$ and computes $X_{SMD_b} = H_{SHA256}(x_{SMD_b})$ and shared session key as $SSK_{SMD_b}^{SMD_a} = H_{SHA384}(y_{SMD_a}, \allowbreak y_{SMD_b}, \allowbreak v_{SMD_b}, \allowbreak v_{SMD_a}, \allowbreak X_{SMD_b})$. Afterwards, it computes key $K_{SMD_b} = H_{SHA256}(k_{SMD_b})$, encrypts required authentication parameters as $E_{SMD_b} = SE^{AES256}_{K_{SMD_b}}(y_{SMD_b}, v_{SMD_b})$, calculates random $R_{SMD_b} = H_{SHA256}(r_{SMD_b})$, encrypts key $K_{SMD_b}$ using the retrieved public key of $SMD_a$ as $C_{SMD_b}^1 = KE_{CPAKE}(PK_{SMD_a}, \allowbreak K_{SMD_b}, \allowbreak R_{SMD_b})$, computes $C_{SMD_b}^2 = KE_{CPAKE}(X_{SMD_a}, \allowbreak X_{SMD_b}, \allowbreak R_{SMD_b})$ and hashed value $H_{SMD_b} = H_{SHA384}(SSK_{SMD_b}^{SMD_a}, \allowbreak y_{SMD_b}, \allowbreak v_{SMD_b}, \allowbreak X_{SMD_b}, \allowbreak T_{SMD_a})$, and returns response $\{E_{SMD_b}, \allowbreak C_{SMD_b}^1, \allowbreak C_{SMD_b}^2, \allowbreak H_{SMD_b}\}$ to $SMD_a$.\\

When $SMD_a$ receives response from $SMD_b$, using its secret key decrypts $C_{SMD_b}^1$ as $K_{SMD_b} = KD_{CPAKE}(SK_{SMD_a}, \allowbreak C_{SMD_b}^1)$ and obtains key $K_{SMD_b}$. Then using $K_{SMD_b}$ decrypts $E_{SMD_b}$ as $(y_{SMD_b}, \allowbreak v_{SMD_b}) = SD^{AES256}_{K_{SMD_b}}(E_{SMD_b})$ and gets $y_{SMD_b}, v_{SMD_b}$. Further, using session-random key $x_{SMD_a}$, decrypts $C_{SMD_b}^2$ as $X_{SMD_b} = KD_{CPAKE}(x_{SMD_a}, C_{SMD_b}^2)$. Having access to these values, it then computes session key as $SSK_{SMD_a}^{SMD_b} = H_{SHA384}(y_{SMD_a}, \allowbreak y_{SMD_b}, \allowbreak v_{SMD_b}, \allowbreak v_{SMD_a}, \allowbreak X_{SMD_b})$. Next, it checks whether $H_{SHA384}(SSK_{SMD_a}^{SMD_b}, \allowbreak y_{SMD_b}, \allowbreak v_{SMD_b}, \allowbreak X_{SMD_b}, \allowbreak T_{SMD_a}) \stackrel{?}{=} H_{SMD_b}$ and if so, it ensures the authenticity and integrity of session key. Following, it sends acknowledgement message $\{A_{SMD_a} = H_{SHA384}(EID_{SMD_a}, SSK_{SMD_a}^{SMD_b}, \allowbreak T_{SMD_a})\}$ to $SMD_b$. Eventually, $SMD_b$ checks $H_{SHA384}(EID_{SMD_a}, \allowbreak SSK_{SMD_b}^{SMD_a}, \allowbreak T_{SMD_a}) \stackrel{?}{=} A_{SMD_a}$ and if so, it also ensures the authenticity and integrity of session key.

\subsection{Smart Metering Device Revocation}
Each saved record in the Hedera blockchain contains an $IsRevoked$ boolean field, which indicates whether or not the SMD is revoked. Initially, this field is set to ``False'' and in case of a secret key leakage for example, the respective SMD can be easily revoked by setting the field value to ``True.'' For this purpose, an owner in the system, which can be a ERA, calls the the ``revokeRecord'' function of the presented smart contract in Fig.~\ref{fig:solidity-smart-contract}. Later on, when a SMD checks the field, it knows that the target SMD is within the revoked ones and stops the process.

\begin{table*}[!htbp]
\setlength{\tabcolsep}{0.3em}
\caption{Designed Protocol}  
\scriptsize 
\label{protocol}
\centering
\begin{threeparttable}
\renewcommand{\arraystretch}{1}
\begin{tabular}{|ll|}
\hline
\rowcolor{LightOrinoco} \multicolumn{2}{|l|}{\textbf{IoT-Enabled Smart Metering Device (SMD) Registration by Semi-Trusted Edge Registration Authority (ERA):}} \\
$SMD_a (SMD_b)$                                                                    &$ERA_c (ERA_d)$ \\
\hline
&\\
&$\overleftarrow{\{Params\}}$\\
$(PK_{SMD_a}, SK_{SMD_a}) = KKG_{CPAKE}()$ &\\
Generates random $v_{SMD_a}$                  &Generates random $x_{ERA_c}$\\
$V_{SMD_a} = H_{SHA256}(v_{SMD_a})$           &$EID_{SMD_a} = SE^{AES256}_{SK_{ERA_c}}(ID_{SMD_a}, x_{ERA_c})$\\
                                              &$IsRevoked_a = False$\\
$\overrightarrow{\{ID_{SMD_a}, PK_{SMD_a}, V_{SMD_a}\}}$ &Stores $(EID_{SMD_a}, PK_{SMD_a}, V_{SMD_a}, IsRevoked_a)$ in the blockchain\\
&\\
                                              &$\overleftarrow{\{EID_{SMD_a}\}}$\\
&\\
\hline    
\rowcolor{LightOrinoco} \multicolumn{2}{|l|}{\textbf{IoT-Enabled Smart Metering Device Peer-to-Peer (P2P) Mutual Authentication Phase:}} \\                                                                      
$SMD_a$ &$SMD_b$ \\
\hline 
$(X_{SMD_a}, x_{SMD_a}) = KKG_{CPAKE}()$&\\
Generates random $k_{SMD_a}, y_{SMD_a}, r_{SMD_a}$ &Checks $|T_{SMD_b} - T_{SMD_a}| \stackrel{?}{<} \bigtriangleup t$\\ 
$K_{SMD_a} = H_{SHA256}(k_{SMD_a})$ &$K_{SMD_a} = KD_{CPAKE}(SK_{SMD_b}, C_{SMD_a})$\\ 
$E_{SMD_a} = SE^{AES256}_{K_{SMD_a}}(EID_{SMD_a}, X_{SMD_a}, y_{SMD_a}, v_{SMD_a})$ &$(EID_{SMD_a}, X_{SMD_a}, y_{SMD_a}, v_{SMD_a}) = SD^{AES256}_{K_{SMD_a}}(E_{SMD_a})$\\ 
Reads $PK_{SMD_b}, V_{SMD_b}, IsRevoked_b$ from the Hedera blockchain &$H_{SHA384}(EID_{SMD_a}, X_{SMD_a}, y_{SMD_a}, v_{SMD_a}, T_{SMD_a}) \stackrel{?}{=} H_{SMD_a}$\\ 
$IsRevoked_b \stackrel{?}{=} False$ &Reads $PK_{SMD_a}, V_{SMD_a}, IsRevoked_a$ from the Hedera blockchain\\ 
$R_{SMD_a} = H_{SHA256}(r_{SMD_a})$ &$H_{SHA256}(v_{SMD_a}) \stackrel{?}{=} V_{SMD_a}$ and $IsRevoked_a \stackrel{?}{=} False$\\
$C_{SMD_a} = KE_{CPAKE}(PK_{SMD_b}, K_{SMD_a}, R_{SMD_a})$ &Generates random $k_{SMD_b}, x_{SMD_b}, y_{SMD_b}, r_{SMD_b}$\\
$H_{SMD_a} = H_{SHA384}(EID_{SMD_a}, X_{SMD_a}, y_{SMD_a}, v_{SMD_a}, T_{SMD_a})$   &$X_{SMD_b} = H_{SHA256}(x_{SMD_b})$\\ 
&$SSK_{SMD_b}^{SMD_a} = H_{SHA384}(y_{SMD_a}, y_{SMD_b}, v_{SMD_b}, v_{SMD_a}, X_{SMD_b})$\\ 
$\overrightarrow{\{E_{SMD_a}, C_{SMD_a}, H_{SMD_a}, T_{SMD_a}\}}$ &$K_{SMD_b} = H_{SHA256}(k_{SMD_b})$\\ 
&$E_{SMD_b} = SE^{AES256}_{K_{SMD_b}}(y_{SMD_b}, v_{SMD_b})$\\
$K_{SMD_b} = KD_{CPAKE}(SK_{SMD_a}, C_{SMD_b}^1)$    &$R_{SMD_b} = H_{SHA256}(r_{SMD_b})$\\                                                                        
$(y_{SMD_b}, v_{SMD_b}) = SD^{AES256}_{K_{SMD_b}}(E_{SMD_b})$ &$C_{SMD_b}^1 = KE_{CPAKE}(PK_{SMD_a}, K_{SMD_b}, R_{SMD_b})$\\
$X_{SMD_b} = KD_{CPAKE}(x_{SMD_a}, C_{SMD_b}^2)$ &$C_{SMD_b}^2 = KE_{CPAKE}(X_{SMD_a}, X_{SMD_b}, R_{SMD_b})$\\
$SSK_{SMD_a}^{SMD_b} = H_{SHA384}(y_{SMD_a}, y_{SMD_b}, v_{SMD_b}, v_{SMD_a}, X_{SMD_b})$ &$H_{SMD_b} = H_{SHA384}(SSK_{SMD_b}^{SMD_a}, y_{SMD_b}, v_{SMD_b}, X_{SMD_b}, T_{SMD_a})$\\
$H_{SHA384}(SSK_{SMD_a}^{SMD_b}, y_{SMD_b}, v_{SMD_b}, X_{SMD_b}, T_{SMD_a}) \stackrel{?}{=} H_{SMD_b}$ &\\
$A_{SMD_a} = H_{SHA384}(EID_{SMD_a}, SSK_{SMD_a}^{SMD_b}, T_{SMD_a})$ &$\overleftarrow{\{E_{SMD_b}, C_{SMD_b}^1, C_{SMD_b}^2, H_{SMD_b}\}}$\\
&\\
$\overrightarrow{\{A_{SMD_a}\}}$ &$H_{SHA384}(EID_{SMD_a}, SSK_{SMD_b}^{SMD_a}, T_{SMD_a}) \stackrel{?}{=} A_{SMD_a}$\\
&\\
\hline
\end{tabular}
Note: The communication channel between SMDs and ERAs during the registration phase is considered safe. 
\end{threeparttable}
\end{table*}

\section{Security Analysis}

\subsection{Concise Informal Verification}
Table \ref{Informal} informally gives key reasons for providing/resisting certain security features/attacks.

\begin{table*}
\setlength{\tabcolsep}{0.1em}
\scriptsize
\centering
\renewcommand{\arraystretch}{1}
\caption{Concise Informal Verification}
\label{Informal}
\begin{threeparttable}
\begin{tabular}{*{3}{l @{\hspace{0.2 cm}} l}}\hline\hline
\rowcolor{LightOrinoco} Property/Attack                  &Reason                   \\\hline
Impersonation Attack     &$C_{SMD_a}$ and $C_{SMD_b}^1$ are two ciphertexts, which can only be decrypted by $SK_{SMD_b}$ and $SK_{SMD_a}$, respectively.\\
Key Compromise Impersonation Attack     &Having keys $SK_{SMD_b}, v_{SMD_b}$, without $SK_{SMD_a}$, attacker cannot impersonate $SMD_a$ and get session key.\\
Replay Attack            &Throughout the protocol, universal time $T_{SMD_a}$ is used in hashed tokens and verified by both $SMD_a$ and $SMD_b$.\\
Unlinkability            &$EID_{SMD_a}$ is encrypted in each session by a fresh key $K_{SMD_a}$ generated by a fresh session-random $k_{SMD_a}$. \\
Perfect Forward Secrecy  &$SSK_{SMD_b}^{SMD_b} = SSK_{SMD_a}^{SMD_b}$ is computed by generating new random $X_{SMD_b}$ in each session.\\
Conditional Anonymity    &Real identifier of $SMD_a/SMD_b$ is encrypted and only distinguishable by $ERA_c/ERA_d$, registered there.\\
Revocability             &Revocation of $SMD_a$ or $SMD_b$ can be done simply by updating the blockchain and setting $IsRevoked = True$.\\
Transparency             &Hedera's blockchain records are transparent and at the disposal of public and can be easily traced and verified.\\
Quantum-Resistance       &The protocol utilizes CRYSTALS-Kyber quantum-secure cryptographic suite and quantum-secure Hedera blockchain.\\
\hline\hline
\end{tabular}
\end{threeparttable}
\end{table*}

\subsection{ProVerif Automatic Formal Verification}
\begin{figure}[!t]
\includegraphics[width=3.5in]{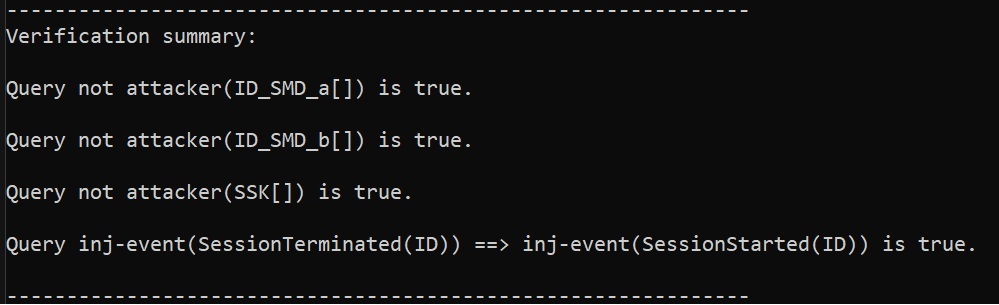}
\caption{Security test results of the ProVerif.}
\label{fig02}
\end{figure}

ProVerif is a powerful and widely used tool for analyzing and proving the security of cryptographic protocols, particularly authentication and key agreement mechanisms \cite{blanchet2013automatic}. Its applicability lies in its ability to model and analyze protocols within a formal framework, enabling the detection of vulnerabilities to replay attacks, impersonation attempts, and message alteration. Through its automated reasoning capabilities, ProVerif efficiently verifies whether a protocol adheres to essential security properties, such as integrity and confidentiality. This makes ProVerif indispensable for designing and validating robust cryptographic protocols, as it provides confidence in their resilience against sophisticated adversarial strategies in real-world applications. The proposed protocol has been modelled and tested using its binary package in windows. Fig.~\ref{fig02} illustrates the security test results of the ProVerif. Full source code can be downloaded from \cite{PS}. As implied, all security goals have been reached demonstrating the secrecy of session key, anonymity of SMDs, and resistance of the protocol to impersonation and replay attacks.

\subsection{Random Oracle Model Formal Verification}
The Random Oracle Model (ROM) plays a crucial role in proving the security of authenticated key agreement protocols by providing a structured framework to analyze potential adversarial strategies \cite{abbasinezhad2018anonymous}. To assess the protocol's security, adversarial interaction is modeled through well-defined queries such as $Execute()$, simulating a passive observer of honest executions; $Send()$, representing active interference by modifying or injecting messages; and $Corrupt()$, enabling the adversary to compromise certain entities. The security proof demonstrates that an adversary's advantage in breaking the protocol—whether by revealing session keys, impersonating a participant, or other attacks—remains negligible under these query-based interactions. This rigorous analysis establishes the resilience of the protocol to passive and active adversaries.

\begin{table}
\setlength{\tabcolsep}{0.1em}
\scriptsize
\centering
\renewcommand{\arraystretch}{1}
\caption{ROM Equations}
\label{equations}
\begin{threeparttable}
\begin{tabular}{l}
\hline\hline
$(3): Adv^{Att_{qe}^{ck}}_{P2PET} = |2Pr(Succ_{subgame_{0}}^{Att_{qe}^{ck}})-1|$\\
$(4): Pr(Succ_{subgame_{1}}^{Att_{qe}^{ck}}) = Pr(Succ_{subgame_{0}}^{Att_{qe}^{ck}})$ \label{3}\\
$(5): |Pr(Succ_{subgame_{2}}^{Att_{qe}^{ck}}) - Pr(Succ_{subgame_{1}}^{Att_{qe}^{ck}})| \leq \frac{(q_{S} + q_{E})^2}{2^{6144}} + \frac{q_{H}^2}{2^{385}}$\\
$(6): |Pr(Succ_{subgame_{3}}^{Att_{qe}^{ck}}) - Pr(Succ_{subgame_{2}}^{Att_{qe}^{ck}})| \leq \frac{q_{S}}{2^{384}}$\\
$(7): |Pr(Succ_{subgame_{4}}^{Att_{qe}^{ck}}) - Pr(Succ_{subgame_{3}}^{Att_{qe}^{ck}})| \leq Adv^{Att_{qe}^{ck}}_{Kyber} + Adv^{Att_{qe}^{ck}}_{Kyber}$\\
$(8): |Pr(Succ_{subgame_{5}}^{Att_{qe}^{ck}}) - Pr(Succ_{subgame_{4}}^{Att_{qe}^{ck}})| \leq Adv^{Att_{qe}^{ck}}_{AES}$\\ 
\hline\hline
\end{tabular}
\end{threeparttable}
\end{table}

\begin{table}
\setlength{\tabcolsep}{0.1em}
\footnotesize
\centering
\renewcommand{\arraystretch}{1}
\caption{Proof of Theorem 1}
\label{theorem}
\begin{threeparttable}
\begin{tabular}{c}
\hline\hline
$Adv^{Att_{qe}^{ck}}_{P2PET} = |2Pr(Succ_{subgame_{0}}^{Att_{qe}^{ck}})-1| =$\\
$2Pr(Succ_{subgame_{5}}^{Att_{qe}^{ck}})-1\; + 2(Pr(Succ_{subgame_{0}}^{Att_{qe}^{ck}})-$\\
$Pr(Succ_{subgame_{5}}^{Att_{qe}^{ck}})) \leq 2Pr(Succ_{subgame_{5}}^{Att_{qe}^{ck}})-1\; +$ \\
$2\sum_{i=0}^4 (Pr(Succ_{subgame_{i+1}}^{Att_{qe}^{ck}}) - Pr(Succ_{subgame_{i}}^{Att_{qe}^{ck}})) = 2 \times \frac{1}{2} -1 +$\\
$2(\frac{(q_{S} + q_{E})^2}{2^{6144}} + \frac{q_{H}^2}{2^{385}} + \frac{q_{S}}{2^{384}} + 2Adv^{Att_{qe}^{ck}}_{Kyber} +$ $Adv^{Att_{qe}^{ck}}_{AES})$\\
\hline\hline
\end{tabular}
\end{threeparttable}
\end{table}

\begin{table*}[t]
\setlength{\tabcolsep}{0.5em}
\caption{Execution Time Comparative Analysis of Related Protocols}
  \label{performance}
  \scriptsize
  \centering
\renewcommand{\arraystretch}{1}
  \begin{threeparttable}
\begin{tabular}{ccccc}
\hline\hline
\rowcolor{LightOrinoco} Client Side Computations and Time
&Server Side Computations and Time
&Total Execution Time 
&Number of Messages
&Protocol\\\hline
$5h + pa + 4pm \approx 2373\, \mu s$   &$5h + pa + 4pm \approx 521\, \mu s$   &$\approx 2894\, \mu s$ &$3$      &\cite{abbasinezhad2018anonymous}\\
$4h + 2ed + 3pm \approx 1782\, \mu s$  &$5h + 2ed + 3pm  \approx 392\, \mu s$ &$\approx 2174\, \mu s$ &$2$      &\cite{kumar2018lightweight}\\
$4h + 2ed + 3pm \approx 1782\, \mu s$  &$6h + 2ed + 4pm \approx 522\, \mu s$  &$\approx 2304\, \mu s$ &$2$      &\cite{chaudhry2022sg}\\
$4h + pa + 4pm \approx 2370\, \mu s$   &$4h + pa + 4pm \approx 520\, \mu s$   &$\approx 2890\, \mu s$ &$2$      &\cite{chaudhry2021privacy}\\
$4h + 2pa + 4pm \approx 2373\, \mu s$  &$4h + 2pa + 4pm \approx 520\, \mu s$  &$\approx 2893\, \mu s$ &$3$      &\cite{abbasinezhad2018design}\\
$5h + pa + 4pm \approx 2373\, \mu s$   &$6h + 2pa + 6pm \approx 780\, \mu s$  &$\approx 3153\, \mu s$ &$3$      &\cite{he2016lightweight}\\
$6h + 3pm + ex \approx 2971\, \mu s$   &$6h+2pm+ex+2bp \approx 1152\, \mu s$  &$\approx 4123\, \mu s$ &$3$      &\cite{odelu2016provably}\\
$4h + pa + 4pm \approx 2370\, \mu s$   &$6h + 2pa + 5pm \approx 651 \, \mu s$ &$\approx 3021\, \mu s$ &$2$      &\cite{qi2020two}\\
$4h + pa + 3pm \approx 1781\, \mu s$   &$4h + pa + 3pm \approx 391\, \mu s$   &$\approx 2172\, \mu s$ &$2$      &\cite{hu2022provably}\\
$2h+2ed+3pm+4ex \approx 6520\,\mu s$   &$2h+2ed+3pm+4ex \approx 1266\, \mu s$ &$\approx 7786\, \mu s$ &$2$      &\cite{xiao2021secure}\\
$3h + 5ex \approx 5939\, \mu s$        &$3h + 5ex \approx 1097\, \mu s$       &$\approx 7036\, \mu s$ &$4$  &\cite{baruah2021authenticated}\\
$7h + 3ed + pa + 2pm \approx 1206\, \mu s$ &$6h + 4ed + 3pa + 4pm + bv \approx 2955\, \mu s$ &$\approx 4161\,\mu s$ &$2$ &\cite{shahidinejad2024highly}\\
$5h + pa + 4pm \approx 2373\, \mu s$   &$6h + pa + 4pm + br \approx 2931\, \mu s$ &$\approx 5304\,\mu s$  &$2$  &\cite{wang2019blockchain}\\

$2h + 4ht + 2ed + kg + ke + 2kd + hr \approx \mathbf{249}\, \mu s$   
&$4h + 4ht + 2ed + 2ke + kd + hr \approx \mathbf{247}\, \mu s$  
&$\approx \mathbf{496}\, \mu s$  
&$3$    
& Ours\\
\hline\hline
\end{tabular}
Note: For the client side, we have considered the obtained times for Raspberry Pi 3 B+ (first column of Table~\ref{time}) and for the server side, we have considered the obtained times for Intel Core-i3 10110U CPU (second column of Table~\ref{time}).
\end{threeparttable}
\end{table*}
\begin{table}[t]
\setlength{\tabcolsep}{0.7em}
\caption{Operation Times of Crypto-Elements}
  \label{time}
  \scriptsize
  \renewcommand{\arraystretch}{0.9}
  \centering
  \begin{threeparttable}
\begin{tabular}{lccc}
\hline\hline
\rowcolor{LightOrinoco}&&\multicolumn{2}{c}{Operation Time}\\
\cline{3-4}
\rowcolor{LightOrinoco} Crypto-Operations                  &Symbol   &On ARMv8                          &On Intel Core-i3 \\\hline
``SHA256'' Hashing Operation       &$h$      &$3\, \mu s$                       &$0.8\, \mu s$\\
``SHA384'' Hashing Operation       &$ht$     &$5.7\, \mu s$                     &$1.5\, \mu s$\\
``AES256'' Encryption/Decryption   &$ed$     &$1.6\, \mu s$                     &$0.6\, \mu s$\\
``nistb163'' Point Multiplication  &$pm$     &$589\, \mu s$                     &$129\, \mu s$\\
``nistb163'' Point Addition        &$pa$     &$2.4\, \mu s$                     &$0.5\, \mu s$\\
``Kyber512'' Key Generation        &$kg$     &$48\, \mu s$                      &8.6$\, \mu s$\\
``Kyber512'' Encryption            &$ke$     &$50\, \mu s$                      &8.4$\, \mu s$\\
``Kyber512'' Decryption            &$kd$     &$9.3\, \mu s$                      &0.7$\, \mu s$\\
``HLF'' Read Operation             &$fr$     &-                                 &2410$\, \mu s$\\
``HLF'' Verification Operation     &$fv$     &-                                 &2430$\, \mu s$\\
``Hedera'' Read Operation          &$hr$     &-                                 &100$\, \mu s$\\
Exponentiation                     &$ex$     &$1186\, \mu s$                    &$219\, \mu s$\\
Bilinear Pairing                   &$bp$     &$1531\, \mu s$                    &$335\, \mu s$\\
\hline\hline
\end{tabular}
HLF: Hyperledger Fabric
\end{threeparttable}
\end{table}

Within the ROM, $Att_{qe}^{ck}$ can interact and send requests to $SMD_a$ and $SMD_b$ using the specified queries in Table~\ref{Queries}. To simulate these interactions in the proposed "P2P Energy Trading (P2PET)" Protocol, game $Game_{P2PET}^{Att_{qe}^{ck}}$ is defined, where $Att_{qe}^{ck}$ can send multiple queries with the main goal of acquiring the session key. However, the final $TestSSK(SMD_a, SMD_b)$ query can only be issued once. Through the following theorem, it will be demonstrated that $Att_{qe}^{ck}$ has only a negligible chance to compromise the security of the proposed P2PET. Formally, assuming $Succ(Game_{P2PET}^{Att_{qe}^{ck}})$ as the success of $Att_{qe}^{ck}$ in the game $Game_{P2PET}^{Att_{qe}^{ck}}$, the advantage of $Att_{qe}^{ck}$ is defined as $Adv^{Att_{qe}^{ck}}_{P2PET}=|2Pr(Succ(Game_{P2PET}^{Att_{qe}^{ck}}))-1| < \epsilon$. Doing so, the security and robustness of P2PET protocol are established.

\textit{Theorem 1}. Assuming $q_{E}$ as the total number of $Execute(SMD_a, SMD_b)$ queries, $q_{H}$ as the total number of $SHA3(Inp)$ queries, $q_{S}$ as the total of $Send(SMD_a, Req)$ or $Send(SMD_b, Req)$ queries, $Adv^{Att_{qe}^{ck}}_{AES}$ as the advantage of $Att_{qe}^{ck}$ against "AES256," and $Adv^{Att_{qe}^{ck}}_{Kyber}$ as the advantage of $Att_{qe}^{ck}$ against "Kyber512," $Adv^{Att_{qe}^{ck}}_{P2PET}$ is defined as~(\ref{1}).

\begin{equation}{}\label{1}
\begin{split}
Adv^{Att_{qe}^{ck}}_{P2PET} \leq \frac{(q_{S} + q_{E})^2}{2^{6143}} + \frac{q_{H}^2 +  2q_{S}}{2^{384}} + \\
+\: 2(Adv^{Att_{qe}^{ck}}_{AES} + 2Adv^{Att_{qe}^{ck}}_{Kyber}) < \epsilon \qquad
\end{split}
\end{equation}

\textit{Proof}. We split $Game_{P2PET}^{Att_{qe}^{ck}}$ into sub-games $subgame_{0}$ to $subgame_{5}$, which are explained below. If $Att_{qe}^{ck}$ wins sub-game $subgame_{x}$, it will be considered a success event and denoted as $Succ_{subgame_{x}}^{Att_{qe}^{ck}}$. Sub-games are defined as follows.

$subgame_{0}$: Since $subgame_{0}$ is indistinguishable from the real attack, $Succ(Game_{P2PET}^{Att_{qe}^{ck}})$ is equal to $Succ_{subgame_{0}}^{Att_{qe}^{ck}}$ and Equation~(3) is derived.

$subgame_{1}$: In $subgame_{1}$, $SMD_a$ and $SMD_b$ instances respond to $Att_{qe}^{ck}$ exactly as specified in the protocol design so Equation~(4) is derived.

$subgame_{2}$: This sub-game is distinguishable from $subgame_{1}$ only when a Hash or transcript collision occurs. Since a Hash collision occurs with a probability of $\frac{q_{H}^2}{2^{385}}$, and a transcript collision with a probability of $\frac{(q_{S} + q_{E})^2}{2^{6144}}$, Equation~(5) is derived.

$subgame_{3}$: This sub-game is distinguishable from $subgame_{2}$ only when $Att_{qe}^{ck}$ obtains the value of integrity tokens without issuing a $SHA3(Inp)$ query. Hence, Equation~(6) is derived.

$subgame_{4}$: This sub-game aims at proving the PFS property and KCI attack resistance of the protocol when $Att_{qe}^{ck}$, by issuing $Corrupt(SMD_a/SMD_b)$ query, gets access to long-term keys of $SMD_a/SMD_b$. $Att_{qe}^{ck}$ cannot distinguish the session key without decrypting $C_{SMD_a}/C_{SMD_b}^1$ (for KCI attck) or $C_{SMD_b}^2$ (for PFS); so Equation~(7) is derived.

$subgame_{5}$: This sub-game aims at proving the privacy-preservation property of the proposed protocol. Since to access the real identifier, $Att_{qe}^{ck}$ must decrypt an AES256 encrypted value, Equation~(8) is derived.

Using Equations (3--8), the theorem is proved as Table \ref{theorem}.

\section{Performance Analysis}

\subsection{Hardware Performance Evaluation}
To ensure an impartial and balanced evaluation of related protocols against the one we recommended, we measured the execution time of various crypto-operations on two hardware platforms using the OpenSSL \cite{OpenSSL} and Kyber libraries \cite{ARMKyber, IntelKyber}. The two hardware platforms utilized for this purpose are the “Raspberry Pi 3 B+” and a laptop computer. The Raspberry Pi 3 B+, equipped with 1GB of RAM and a 64-bit ARMv8 Cortex-A53 BCM2837B0 CPU, is used for computing the SMD (client) side timings. Meanwhile, the laptop, featuring 8GB of RAM and an Intel Core-i3 10110U CPU, handles the server side computations. The times for the cryptographic operations on both platforms are summarized in Table~\ref{time}.

\subsection{Comparative Performance Evaluation}
According to the obtained times in Table~\ref{time}, we have computed the client side, server side, and total execution times of top-related protocols together with the proposed one. Table~\ref{performance} demonstrates the results. We have used same hardware platforms for the all schemes to keep conformity and fairness. In Table~\ref{performance}, for the client side, we have considered the Raspberry Pi 3 B+ and for the server side, we have considered the Intel Core-i3 10110U CPU. However, because in the proposed protocol both client and server are smart meters, we have considered the Raspberry Pi 3 B+ for the both sides. According to the obtained results, the initiator SMD and target SMD, respectively, spend almost $249\, \mu s$ and $247\, \mu s$, far better than most related ones. This is while, according to Table~\ref{tab:related}, the proposed protocol provides more features than all related ones, such as transparency, revocability, and quantum-resilient security through the use of CRYSTALS-Kyber, a post-quantum cryptographic algorithm designed to withstand known quantum attacks. The quantum-resilient security of the proposed protocol was further supported through an informal security analysis, a formal security proof under the Random Oracle Model, and automated verification using ProVerif. It should be clarified that, like all other schemes, the proposed protocol has also limitation. Its limitation arises from the inherent limitation of post-quantum techniques such as CRYSTALS-Kyber. The limitation is size of ciphertext that is 768 Bytes. To address this limitation and save bandwidth, data should be compressed before submission, which adds some time to the protocol execution.

\section{Conclusion}
Rapid advancements in quantum computing present profound challenges to the security of peer-to-peer (P2P) energy trading systems, necessitating the development of innovative solutions. This paper has introduced a groundbreaking authenticated session key agreement protocol tailored for edge-IoT P2P energy trading communications, addressing the unique demands of the quantum era. By combining CRYSTALS-Kyber, one of the NIST-approved quantum-resistant cryptographic suites with the quantum-resistant Hedera Hashgraph, the protocol ensures robust security, transparency, and revocability. Extensive analyses affirm its superiority in both security and efficiency, highlighting its minimal computational overhead on resource-constrained IoT-enabled smart metering devices. Through these advancements, the proposed protocol contributes significantly to the creation of resilient and sustainable decentralized energy markets, laying the groundwork for secure and efficient energy distribution in a future shaped by quantum technology.

Future work may focus on enhancing the physical and operational security of IoT-enabled smart meters by incorporating countermeasures against side-channel and physical tampering attacks, which remain a critical concern in real-world deployments. In addition, integrating machine learning-based anomaly detection techniques could provide an adaptive security layer for identifying unusual or malicious trading behaviours in P2P energy markets, thereby improving the system’s resilience against both cyber and behavioural attacks.

\bibliographystyle{IEEEtraN}
\bibliography{references}

\end{document}